\documentclass[12pt,pre,aps,showpacs,preprint]{revtex4}

\usepackage{graphicx}
\usepackage{color}
\usepackage{amsfonts}
\usepackage{amssymb}
\usepackage{amsmath}

\begin{document}

\title{Dense Periodic Packings of Tori}

\author{Ruggero Gabbrielli}
\email{ruggero.gabbrielli@unitn.it} \affiliation{Interdisciplinary
Laboratory for Computational Science, Department of Physics,
University of Trento, 38123 Trento, Italy}

\author{Yang Jiao}
\email{yang.jiao.2@asu.edu}
\affiliation{Materials Science and Engineering Program, School for
Engineering of Matter, Transport and Energy, Arizona State
University, Tempe, Arizona 85281, USA}

\author{Salvatore Torquato}
\email{torquato@electron.princeton.edu}
\affiliation{Department of Chemistry, Department of Physics,}
\affiliation{Program in Applied and Computational Mathematics,}
\affiliation{Princeton Institute for the Science and Technology of Materials,
Princeton University, Princeton, New Jersey 08544, USA}

\date{\today}

\begin{abstract}
Dense packings of nonoverlapping bodies in three-dimensional
Euclidean space $\mathbb{R}^3$ are useful models of the structure
of a variety of many-particle systems that arise in the physical
and biological sciences. Here we investigate the packing behavior
of congruent ring tori in $\mathbb{R}^3$, which are multiply
connected non-convex bodies of genus one, as well as horn and
spindle tori. Specifically, we analytically construct a family of
dense periodic packings of unlinked, congruent tori guided by the
organizing principles originally devised for simply connected
solid bodies [Torquato and Jiao, Phys. Rev. E {\bf 86}, 011102
(2012)]. We find that the horn tori as well as certain spindle and
ring tori can achieve a packing density not only higher than that
of spheres (i.e., $\pi/\sqrt{18}=0.7404\ldots$) but also higher
than the densest known ellipsoid packings (i.e., $0.7707\ldots$).
In addition, we study dense packings of clusters of pair-linked
ring tori (i.e., Hopf links), which can possess much higher
densities than corresponding packings consisting of unlinked tori.

\end{abstract}



\maketitle


\section{Introduction}
\label{intro}

Dense packings of nonoverlapping bodies (i.e., hard particles) are
useful models of a variety of low-temperature state of matter
\cite{Ber65, Za83, Ch00, To02, Pond.JCP.11, Mare.JCP.11}, granular
media \cite{Ed94, To02}, heterogeneous materials \cite{To02}, and
biological systems \cite{Li01, Pu03, polymer}. Probing the
symmetries and other mathematical properties of the densest
packings is a problem of interest in discrete geometry and number
theory \cite{ConwayBook, Henry}. In general, a packing is defined
as a large collection of nonoverlapping solid objects in
$d$-dimensional Euclidean space $\mathbb{R}^d$. Associated with a
packing is the packing fraction (or density) $\phi$ defined as the
fraction of space $\mathbb{R}^d$ covered by the particles.

The densest packings of nonoverlapping objects are of particular
interest because of their relationship to  ground states of matter
and because such configurations determine the high-density phases
in the equilibrium phase diagram of hard-particle systems
\cite{To10}. The determination of the densest packing arrangements
of particles that do not fill all of space \cite{Ga12} is a
notoriously challenging problem. For example, the rigorous proof
of the conjecture for the densest arrangements of congruent (i.e.,
equal-sized) spheres put forth by Kepler \cite{kepler}, the most
elementary non-space-filling shape, took almost four centuries.
The conjecture, now a theorem, states that the packing fraction of
spheres in three-dimensional Euclidean space can not exceed the
value $\phi=\pi/\sqrt{18}= 0.7404\ldots$ \cite{hales-kepler}.
There has been recent progress on the determination of the densest
known packings of spheres of two different sizes \cite{Ho12}, but
there are no rigorous proofs that any of these packings are indeed
the densest for a given size ratio and relative concentration.

The preponderance of studies of dense packings of nonspherical
shapes in $\mathbb{R}^3$ have dealt with convex bodies, which are
always simply connected and thus topologically equivalent to a
sphere. Examples of such convex shapes, which are of genus zero,
include ellipsoids \cite{donev}, superballs \cite{superball,
superball2}, cones \cite{cone}, spherocylinders \cite{organizing},
lens-shaped particles \cite{organizing}, and various polyhedra
\cite{chen,betke,torquato-platonic,jiao,kallus,damasceno-predictive,
degraaf}. More recently, researchers have found dense packings of
non-convex (or concave), simply connected bodies, such as
non-convex superballs (including three-dimensional crosses)
\cite{superball}, dumbbell-shaped particles \cite{dumbbell},
certain non-convex polyhedra (including Szilassi polyhedra,
tetrapods, octapods, and stellated dodecahedra) \cite{degraaf},
and non-convex building blocks composed of clusters of
nonspherical shapes \cite{organizing}. Organizing principles and
conjectures have been proposed that enable one to predict dense
packing arrangements of simply connected bodies based on the
characteristics of the particle shapes (e.g., symmetry, principal
axes, and convexity) \cite{organizing}. This includes a
Kepler-like conjecture for the densest packings of the
centrally-symmetric Platonic and Archimedean solids
\cite{torquato-platonic}. However, much less is known about dense
packings of multiply connected solid bodies.


The focus of this paper is the determination of dense packings of
congruent tori. A ring torus is an elementary multiply connected
non-convex non-space-filling body of genus one, which constitutes
an infinite variety of shapes \cite{fn1}. The interest in such a
family of shapes is not merely theoretical but has practical
applications. One relevant example is in pharmaceutical
technologies. Toroidal particles have been and are currently being
investigated for use in drug delivery as they possess a higher
surface area per unit volume than commonly used simply connected
shapes such as short cylindrical-like pills \cite{ungphaiboon}.
Fabrication methods have proven successful either via colloidal
crystallization \cite{velev} or microfluidics \cite{wang}.

\begin{figure}[h]
\includegraphics[width=0.6\linewidth]{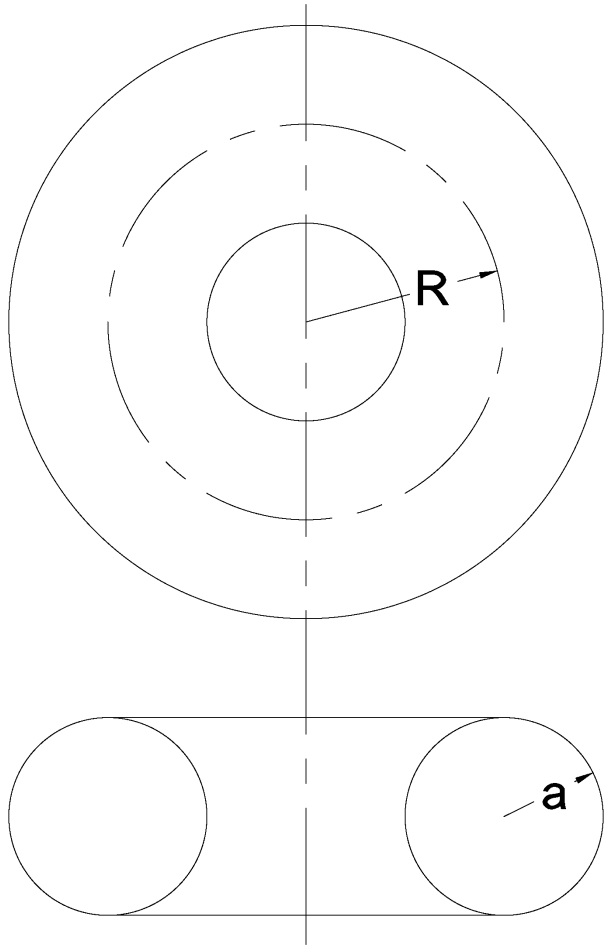}
\caption{\label{torus}The two radii defining a torus: major radius $R$ and minor
radius $a$. We denote the major-to-minor radii ratio by $\gamma=R/a$.}
\end{figure}

\begin{figure}[h]
\includegraphics[width=\linewidth]{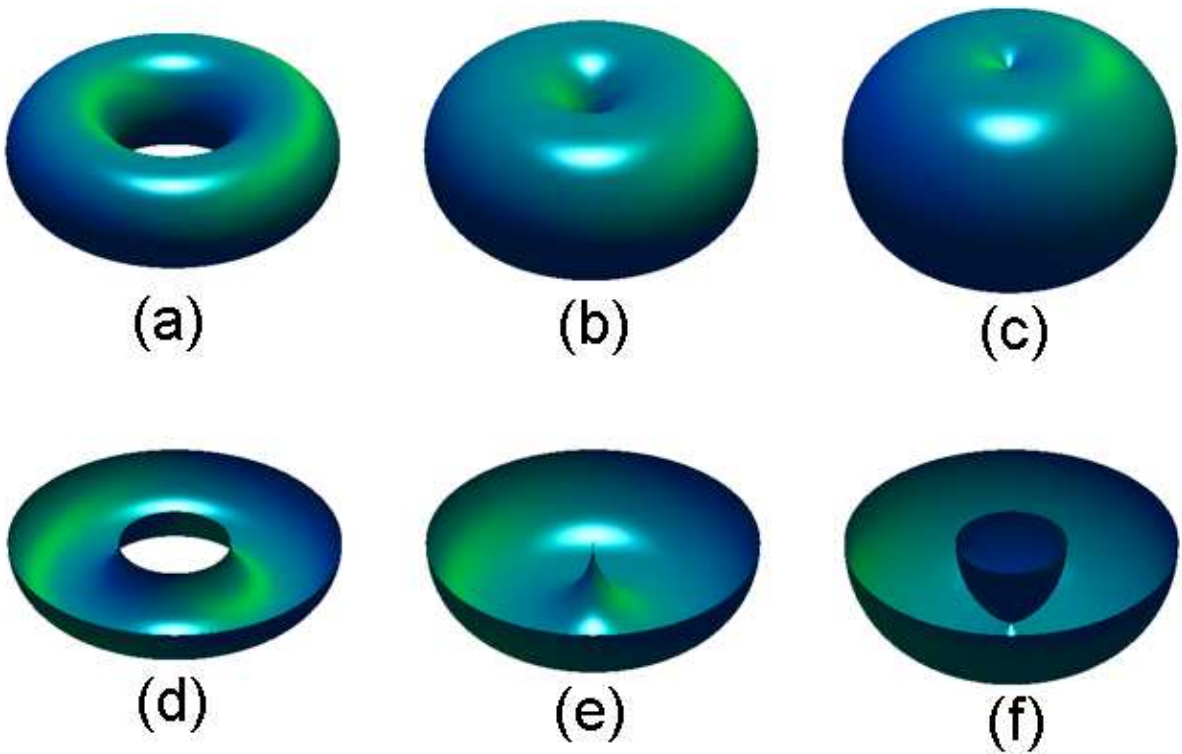}
\caption{\label{tori}Three different types of tori (upper panels) and the
associated cut-aways that divide the tori in half (lower panels)
as classified based on their radii ratio $\gamma$. (a) A ring
torus with $\gamma>1$. (b) A horn torus with $\gamma = 1$. (c) A
spindle torus with $0<\gamma<1$. This is also the surface of an apple.
(d) A cut-away of the ring
torus shown in (a). (e) A cut-away of the horn torus shown in
(b). (f) A cut-away of the spindle torus shown in (c). A half of a ``lemon" is visible within.}
\end{figure}

A torus in $\mathbb{R}^3$ is defined by the surface of revolution
generated by revolving a circle in three dimensions about an axis
that is coplanar with the circle (see Fig.~\ref{torus}). The surface of a
torus centered at the origin can be written in parametric form as:
\begin{equation}
\begin{cases}
x=(R+a \cos\phi) \cos \theta \\
y=(R+a \cos\phi) \sin\theta \\
z= a \sin\phi
\end{cases}
\end{equation}
where $0<\phi,\theta \le 2\pi$ and $R$ is the distance from the
center of the generator circle to the axis of revolution or the
{\it major radius} and $a$ is the radius of the circle itself or
the {\it minor radius}. A torus can take a continuous range of
shapes and is parametrized by the ratio
\begin{equation}
\gamma=\frac{R}{a},
\end{equation}
where $R$ is the major radius and $a$ is the minor radius. In
particular, tori are classified into three different types based
on the value of their major-to-minor radii ratio (or radii ratio
for short). For $\gamma<1$, the surface self-intersects at two
points, thus generating a {\it spindle} torus. By construction,
there is a region inside a spindle torus whose volume should be formally counted twice. This region has the shape of an American football and it is called a \emph{lemon} [Fig.~\ref{tori}(f)]. In this work we only consider the external surface of the spindle torus, also known as the \emph{apple} [Fig.~\ref{tori}(c)]. For $\gamma=1$ the
torus surface is tangent to the torus axis. The surface normal is
defined everywhere apart from a single point, the torus center.
This is called a {\it horn} torus. For $\gamma>1$, the solid body is a
{\it ring} torus. The surface normal is defined everywhere on the
manifolds in such cases. Figure \ref{tori} shows these three different types of tori and
their corresponding cut-aways that divide the tori in half.

If we substitute apples for spindle tori, the
volume $v_{\mbox{\tiny T}}$
and surface area $s_{\mbox{\tiny T}}$
of such bodies are given by
\begin{equation}
v_{\mbox{\tiny T}}=\begin{cases} 2\pi [\frac{(2+\gamma^{2})}{3}\sqrt{1-\gamma^2}+
\gamma
\cos^{-1}(-\gamma)]
\cdot a^3, &
0<\gamma<1, \\
2\pi^2 \gamma \cdot a^3, & \gamma \ge 1,
\end{cases}
\end{equation}
and
\begin{equation}
s_{\mbox{\tiny T}}=\begin{cases} 4\pi
[\sqrt{1-\gamma^2}+\gamma\cos^{-1}(-\gamma)]
\cdot a^2, &
0<\gamma<1, \\
4\pi^2 \gamma \cdot a^2, & \gamma \ge 1.
\end{cases}
\end{equation}
The packing fraction $\phi$ of a configuration at number density
$\rho$ (number of tori per unit volume) and lattice vectors ${\bf a}_i$ is simply given by
\begin{equation}
\phi = \rho v_{\mbox{\tiny T}}.
\end{equation}

In this paper, we analytically construct a family of dense
periodic packings of unlinked ring tori characterized by their
major-to-minor radii ratio $\gamma$. Such constructions are
achieved by generalizing the organizing principles originally
devised for simply connected nonspherical particles (both convex
and non-convex) \cite{organizing}. We find that for the horn tori
($\gamma = 1$), certain spindle ($\gamma<1$) and ring tori
($\gamma > 1$), packings can be achieved with a density $\phi$ not
only higher than that of spheres (i.e., $\pi/\sqrt{18} =
0.7404\ldots$) but also higher than the densest known ellipsoid
packings (i.e., $0.7707\ldots$). In addition, we study dense
packings of clusters of certain linked tori, e.g., dimers composed
of two tori with $\gamma = 2$ linked in the simplest known form
(i.e., a Hopf link).

The rest of the paper is organized as follows: In Sec. II, we
present our analytical construction of a family of dense packings
of tori with different radii ratios. In Sec. III, we construct
dense packings of clusters of linked tori. In Sec. IV, we make
concluding remarks.

\section{Analytical Construction of A Family of Dense Periodic Packings of Tori}

The general organizing principles proposed for simply connected
solid objects suggest that when a centrally symmetric object is
arranged on a {\it Bravais lattice} (one particle per fundamental
cell), the number of contacts that the particle makes with its
neighbors can be maximized, which leads to a dense packing
\cite{organizing}. If the object itself does not possess central
symmetry, usually a centrally symmetric cluster composed of an
even number of the individual objects can be constructed, whose
Bravais-lattice packing provides a dense packing of the solid.
However, the Bravais-lattice packing of the clusters are now a
(non-Bravais lattice) {\it periodic} packing of the original solid
object, since there are multiple objects per fundamental cell.
Since a torus is a centrally symmetric object, we first consider
Bravais lattice packings of unlinked tori.

\begin{figure}[h]
$\begin{array}{c@{\hspace{0.5cm}}c@{\hspace{0.25cm}}c}
\includegraphics[height=2.75cm]{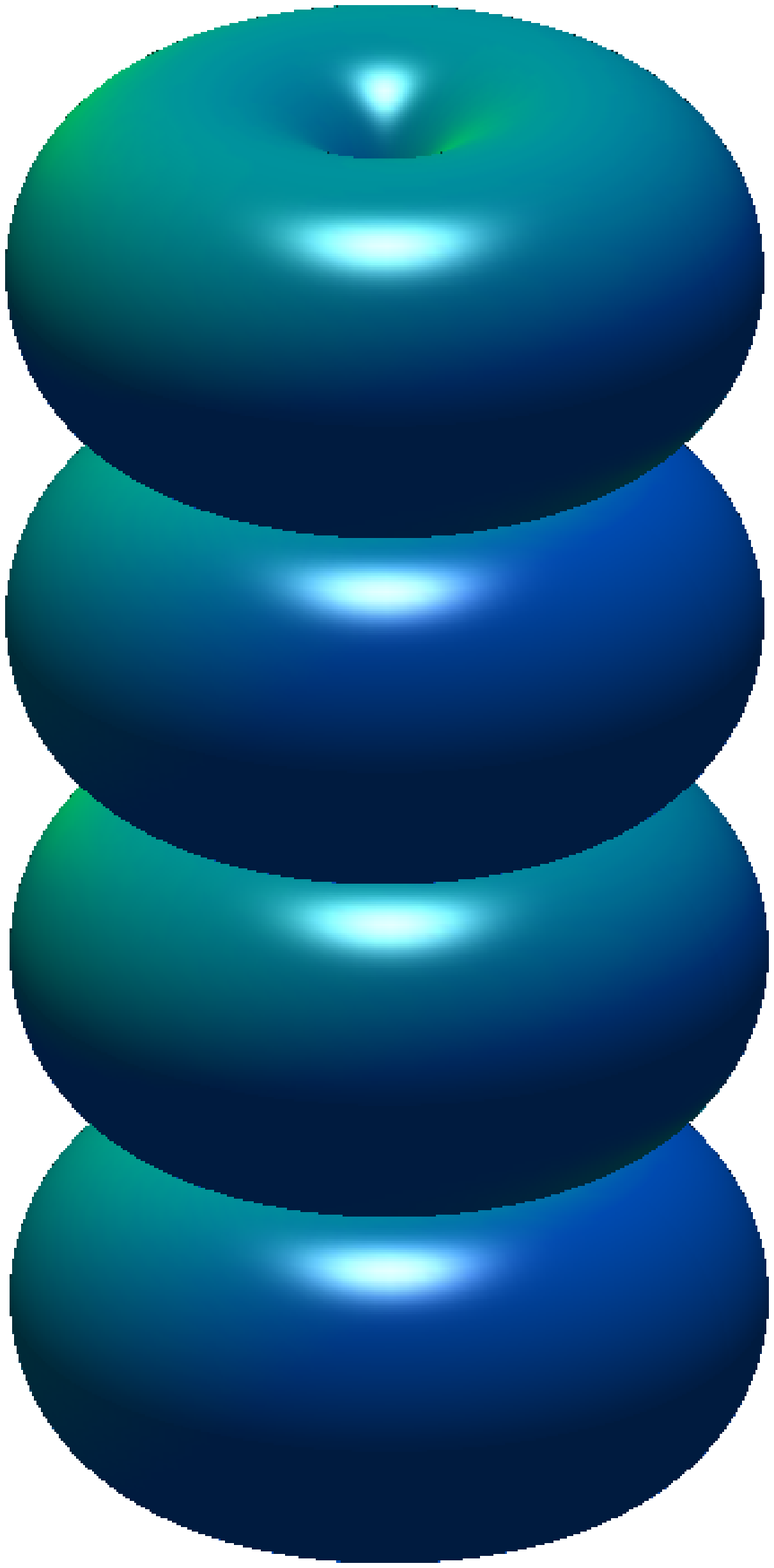} &
\includegraphics[height=2.75cm]{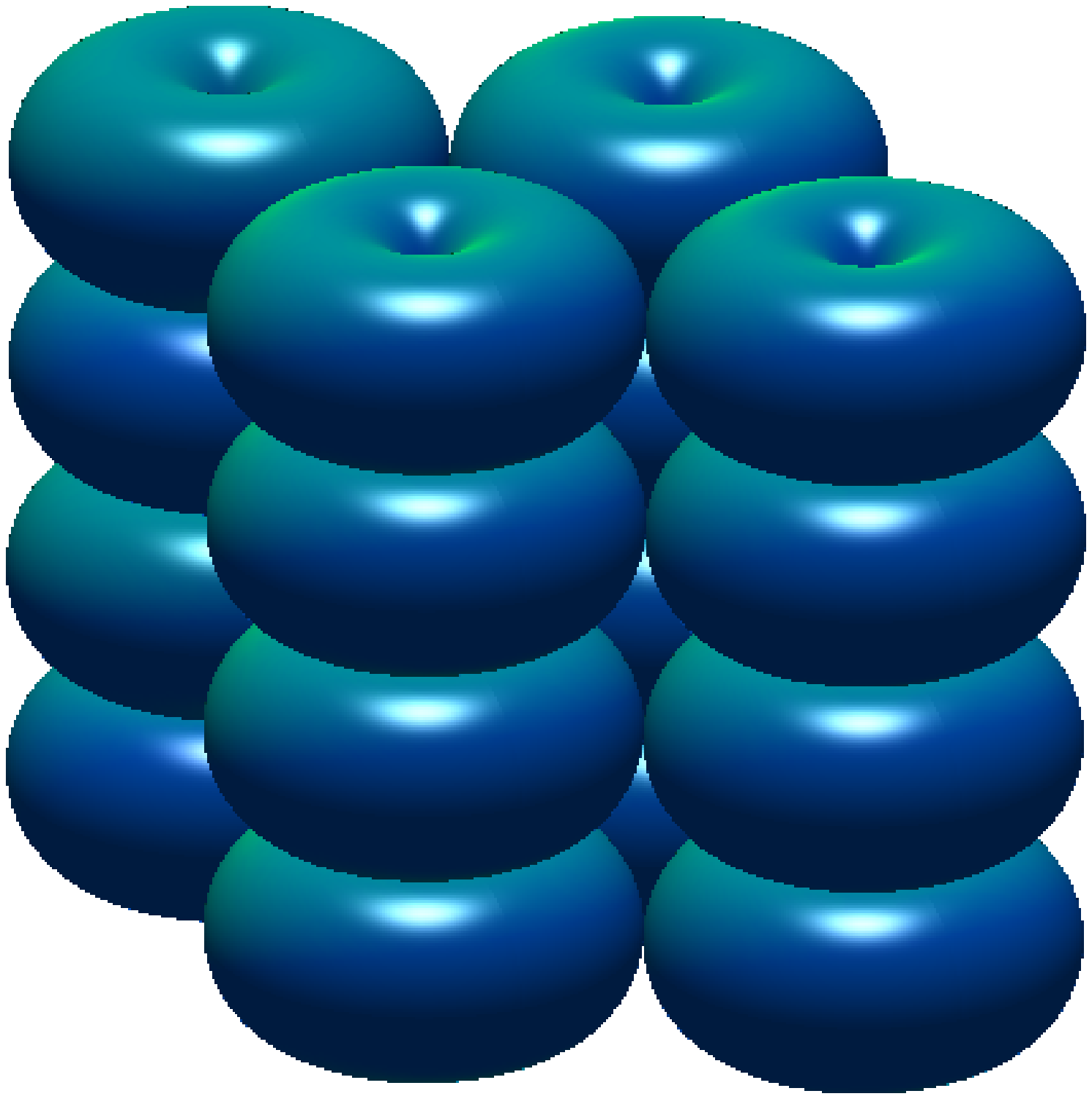} &
\includegraphics[height=2.75cm]{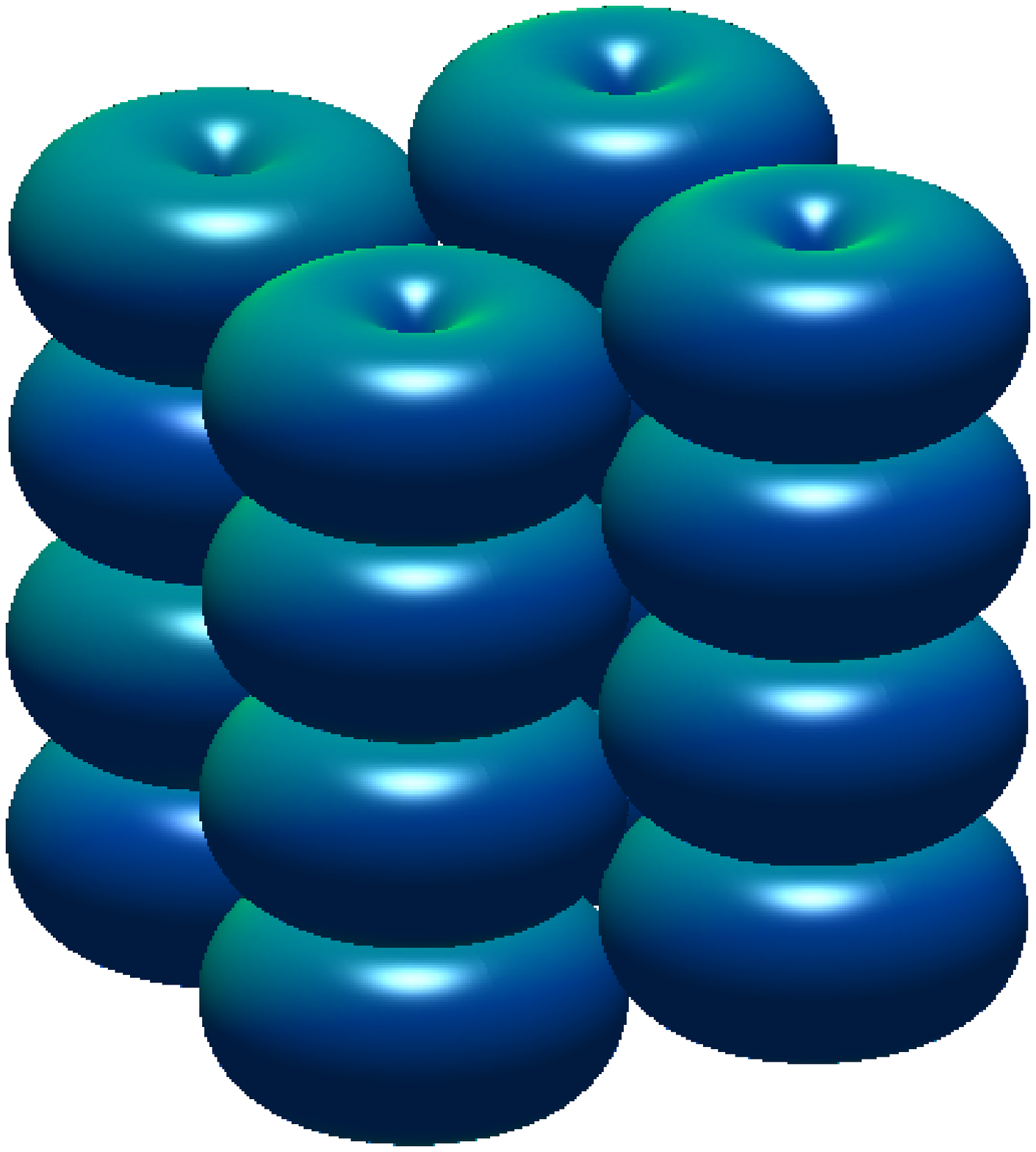} \\
\mbox{\bf (a)} & \mbox{\bf (b)} & \mbox{\bf (c)}\\
\end{array}$
\caption{\label{toripacking}Illustrations of dense periodic packings of tori. (a) A
cylindrical-like cluster constructed by stacking tori exactly on
top of one another. (b) A triangular-lattice packing of the
cylindrical-like clusters of tori in which each torus has 8
contacting neighbors. (c) A dense periodic packing that improves
upon the one shown in (b) with 12 contacting neighbors per torus.}
\end{figure}

Vertically stacking tori exactly on top of each other to make
infinitely large ``cylindrical-like'' clusters, each of which are
then placed on the sites of a triangular lattice, is an
intuitively appealing configuration that one may wish to start
with. In any cylindrical-like cluster, each torus contacts two
neighbors, one above and the other below. Two tori contact one
another through a common circle [see Fig.~\ref{toripacking}(a)].
When the cylindrical-like clusters are arranged on a triangular
lattice, each torus in such a configuration is in contact with 8
neighbors, two of which share a circle and the remaining six are
in contact with a single point lying on the equator [see
Fig.~\ref{toripacking}(b)].

\begin{figure}[h]
\includegraphics[width=\linewidth]{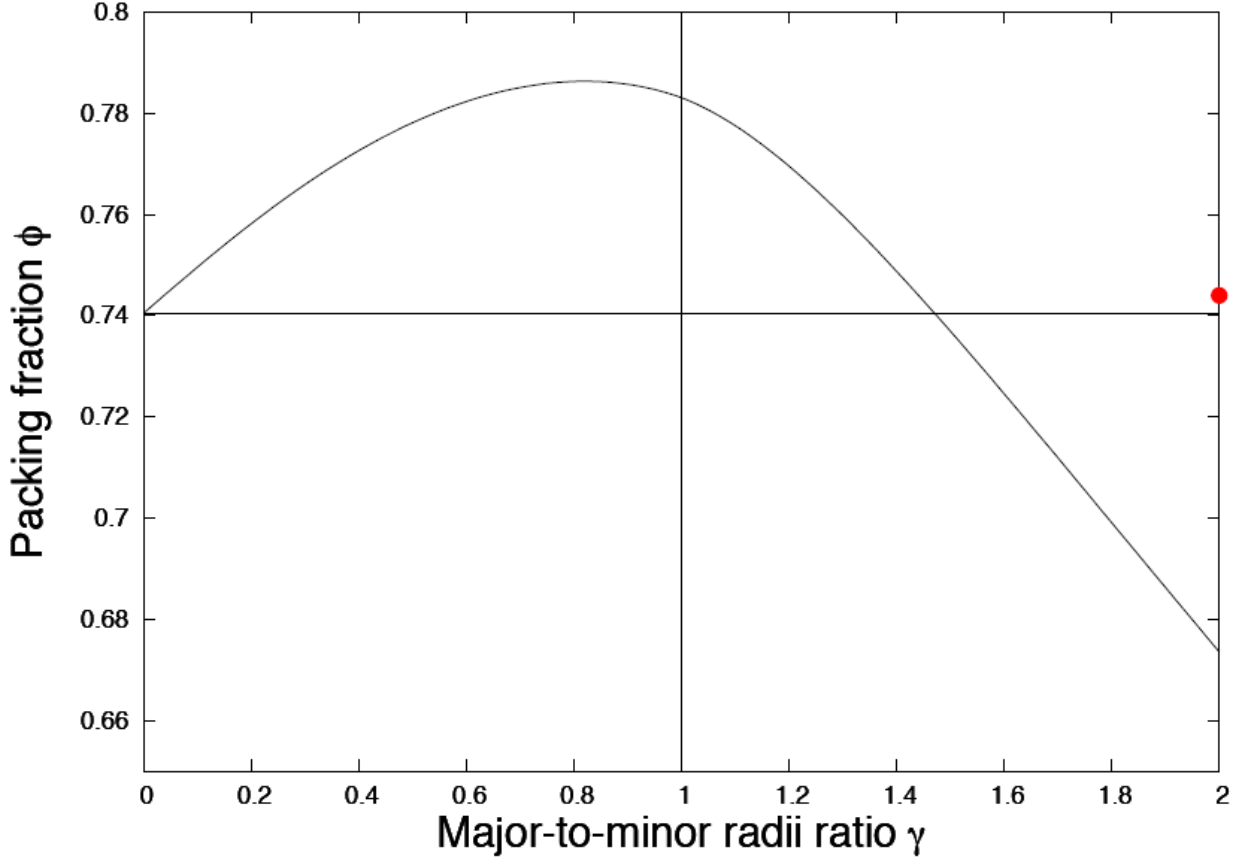}
\caption{\label{packfrac}Packing fraction $\phi$ of the periodic
packing of tori shown in Fig.~2 as a function of the
major-to-minor radii ratio $\gamma$. The packing fraction of the
arrangement in Fig.~\ref{hopf} is also reported at the extreme
value of $\gamma=2$ indicated in the graph as a filled-in circle.}
\end{figure}

We note that a denser packing is obtained by slightly shifting the
cylindrical-like clusters in the triangular-lattice packing along
the axis of the ``cylinders'' so that each torus comes into
contact with four more neighbors (relative to the simple
triangular lattice configuration), thus reaching a ``contact
number'' equal to 12 [see Fig.~\ref{toripacking}(c)]. (Note that two contacts are
also circles in this case.) The lattice vectors associated with
the constructed packing for congruent tori characterized by $R$
and $a$ are given by:
\begin{equation}
\begin{array}{c}
{\bf a}_1 = \left [{\gamma+1, ~\Gamma, ~1}\right ]a,\\
{\bf a}_2 = \left [{-(\gamma+1), ~\Gamma, ~1}\right ]a,\\
{\bf a}_3 = \left [{0, ~0, ~2}\right ]a,
\end{array}
\label{latt1}
\end{equation}
where:
\begin{equation}
\Gamma=\sqrt{3\gamma^2+2(2\sqrt{3}-1)\gamma+2}.
\end{equation}
The associated packing density is given by:
\begin{equation}
\phi(\gamma)=\begin{cases}
\displaystyle{\pi\frac{\gamma \cos^{-1}(-\gamma)+\frac{(2+\gamma^2)\sqrt{1-\gamma^2}}{3}}{2(1+\gamma)\Gamma}} & 0<\gamma<1 \\
\displaystyle{\frac{\pi^2\gamma}{2(1+\gamma)\Gamma}} & \gamma\ge1
\end{cases}
\end{equation}
where $\gamma = R/a$ is the major-to-minor radii ratio of the
tori. Fig.~\ref{packfrac} shows the packing density as a function
of radii ratio $\gamma$. In the case $\gamma = 0$, the torus
reduces to a sphere, which has the optimal packing density of
$\pi/\sqrt{18}=0.7404\ldots$. As $\gamma$ increases from 0, an
increase of packing density $\phi$ is immediately achieved. As the
radii ratio increases, the packing density eventually begins to
decrease due to the emergence of a large void region in the middle
of the cylindrical-like clusters. Nonetheless, a wide spectrum of
tori with radii ratio $\gamma \in (0, ~1.47074\ldots)$ can pack
more densely than spheres in this construction.

Interestingly, a maximum packing density $\phi_{\mbox{\tiny max}}
= 0.786322\ldots$ is achieved by spindle tori of radii ratio
$\gamma^*=0.820265\ldots$. Note that a spindle torus has no holes
and thus is a better packer than the ring tori. The density
associated with the horn tori (with $\gamma = 1$) is given by
\begin{equation}
\phi(\gamma)=\phi(1)=\frac{\pi^2}{4\sqrt{3+4\sqrt{3}}} =
0.783076\ldots
\end{equation}
For ring tori, the maximum density becomes arbitrarily close to
that of horn tori as the radii ratio approaches unity, i.e.,
$\gamma \rightarrow 1$.


\section{Dense Packings of Pair-Linked Tori}

As the major-to-minor radii ratio $\gamma$ becomes large, the hole
in a torus grows in size and makes the cylindrical-like
arrangements inefficient packings. If we allow the tori to be
linked (knotted) with one another, the void space in the torus
hole can be used to improve the packing density. Mathematically,
allowing the tori to link to determine the densest packings of
tori among all possible configurations is perfectly valid given
the fact that a torus has genus one.


First, we note that ring tori with $\gamma \ge 2$ can be assembled
into Hopf links. A Hopf link here refers to pair-linked tori that
are arranged to form a periodic. Then such a ``dimer'' can pair
with an identical object to form a unit made of 4 tori; see Fig.
5. Based on our general organizing principles, a dense of packing
of such a centrally symmetric 4-tori cluster can be achieved by a
Bravais-lattice packing of the clusters, which we describe below.


\begin{figure}[h]
\includegraphics[width=0.8\linewidth]{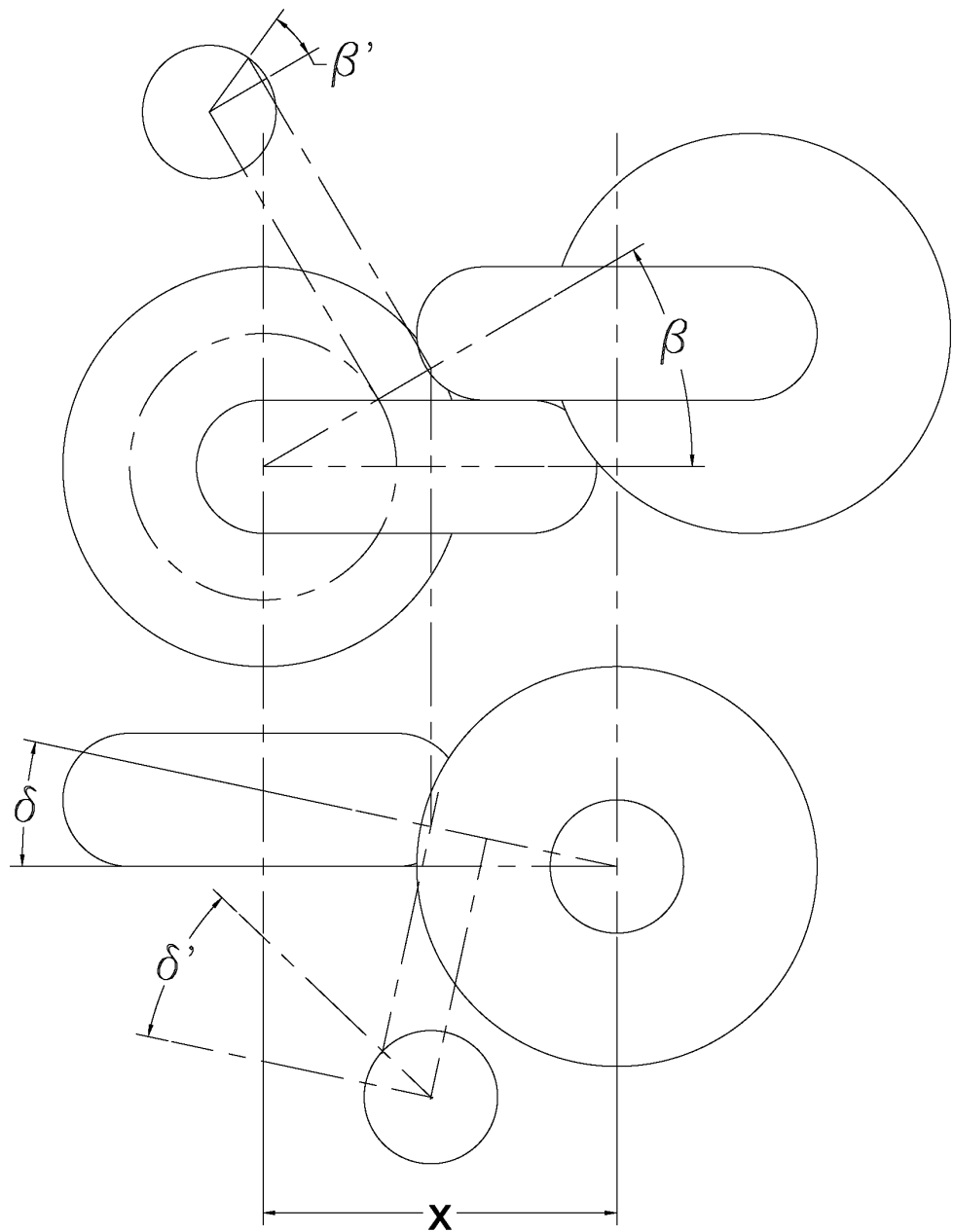}
\caption{\label{hopf-sketch} Contact point between dimers in
the densest known packing of tori of radii
ratio 2. This sketch has been used
for the determination of the only non-trivial lattice vector.}
\end{figure}

\begin{figure}[h]
$\begin{array}{c@{\hspace{0.25cm}}c@{\hspace{0.25cm}}c}
\includegraphics[width=2.65cm]{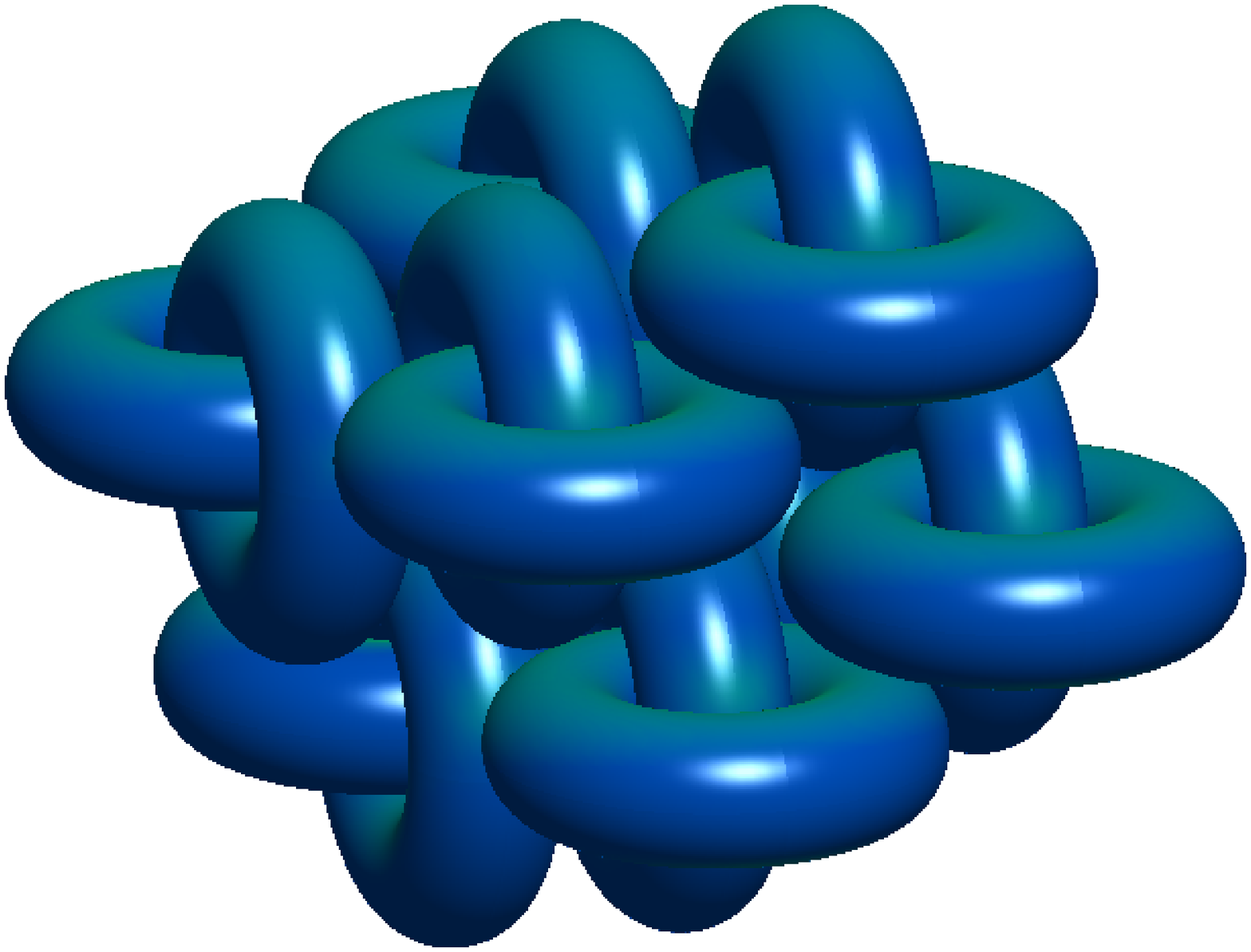} &
\includegraphics[width=2.65cm]{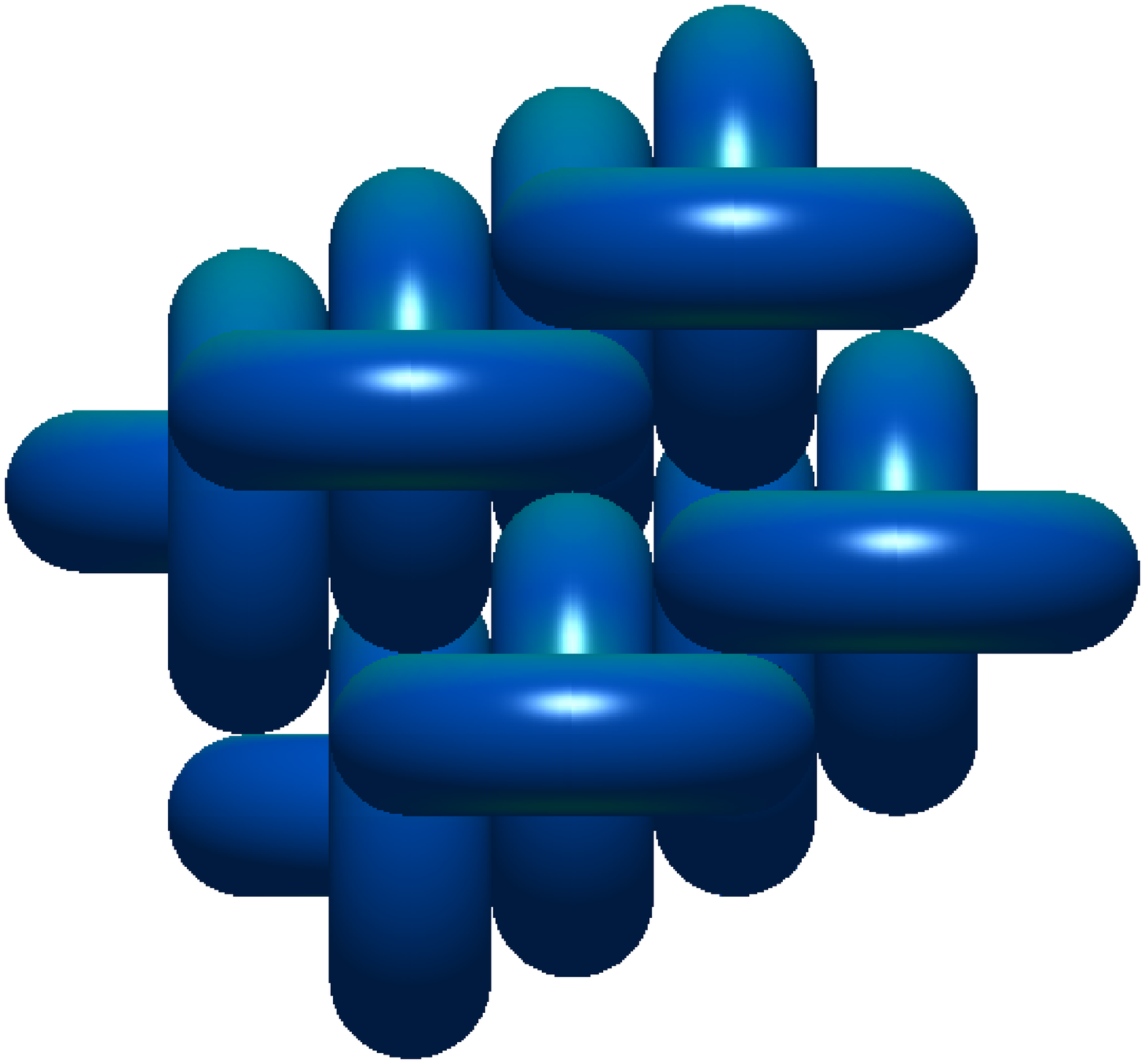} &
\includegraphics[width=2.65cm]{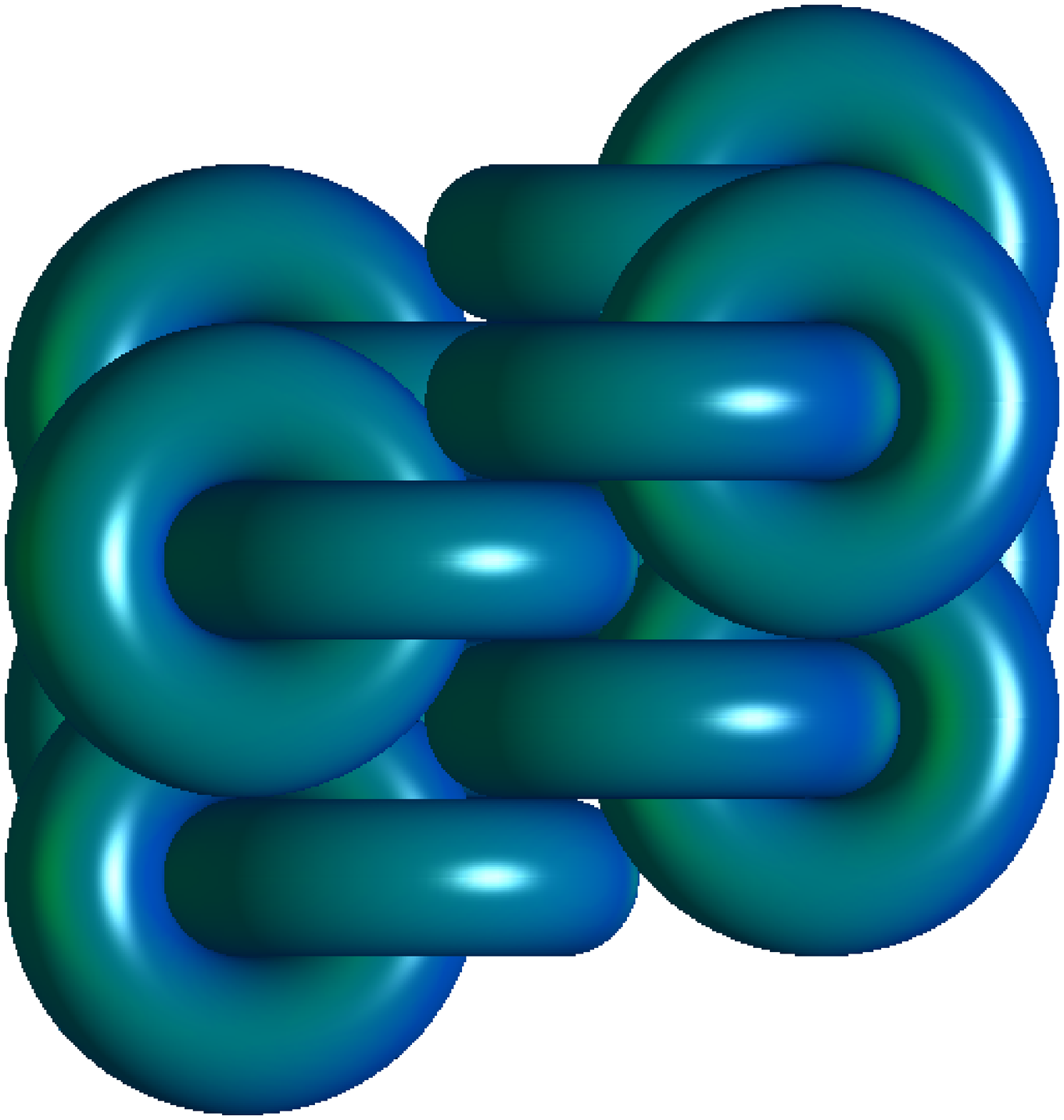} \\
\end{array}$
\caption{\label{hopf} The densest known packing of tori of radii
ratio 2. Packing fraction is $\phi =  0.7445\ldots$. The images
show four periodic units, each containing four tori.}
\end{figure}

Different views of this packing arrangement are shown in Fig.
\ref{hopf}. Lattice vectors and packing fraction can be computed
by solving the following system of transcendental equations:
\begin{equation}
\begin{cases}
\displaystyle{\sin\beta=\frac{2-\sin\delta^{'}}{2+\cos\beta^{'}}} \\
\displaystyle{\sin\delta=\frac{1-\sin\beta^{'}}{2+\cos\delta^{'}}} \\
\displaystyle{\cos^2\beta=\frac{\beta^{'}-1}{\delta-1}} \\
\displaystyle{\cos^2\delta=\frac{\delta^{'}-1}{\beta-1}} \\
x=\cos\beta(2+\cos\beta^{'})+\cos\delta(2+\cos\delta^{'})\\
\end{cases}
\label{syst}
\end{equation}
which yields $\beta = 0.5308\ldots$, $\beta' = 0.4134\ldots$,
$\delta = 0.2113\ldots$, $\delta' = 0.5514\ldots$ and $x =
5.3028\ldots$. The corresponding lattice vectors are given by

\begin{equation}
\begin{array}{c}
{\bf a}_1 = \left [{4, ~2, ~0}\right ]a,\\
{\bf a}_2 = \left [{-2, ~4, ~0}\right ]a,\\
{\bf a}_3 = \left [{1, ~3, ~2x}\right ]a,
\end{array}
\label{latt2}
\end{equation}

and packing fraction is $\phi = 0.7445\ldots$. This significantly
improves up on the density associated with the periodic packing of
unlinked tori based on cylindrical-like clusters and is even
higher than the optimal sphere packing fraction.

\begin{figure}[h]
$\begin{array}{c@{\hspace{0.25cm}}c}
\includegraphics[width=3.75cm]{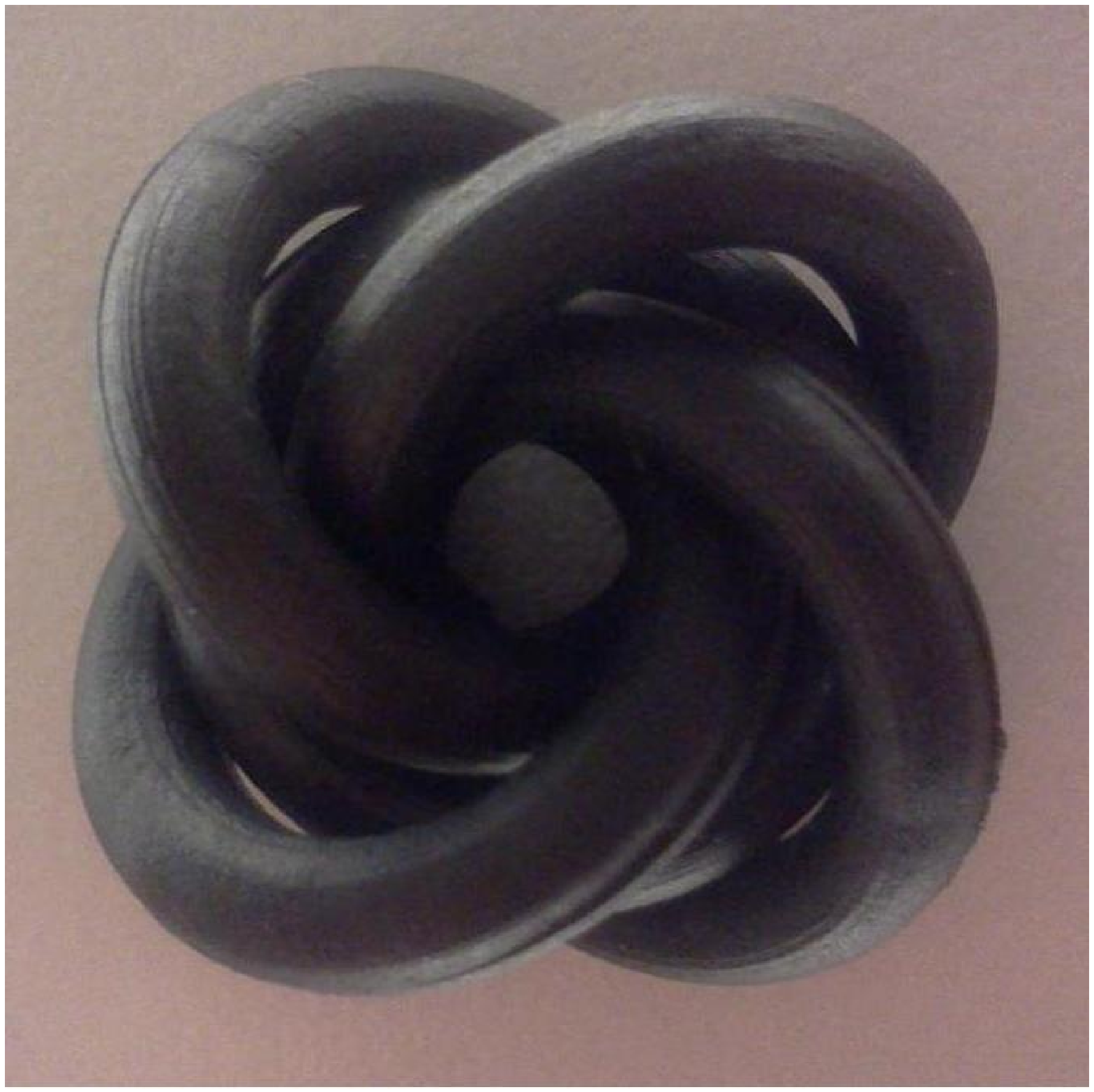} &
\includegraphics[width=3.75cm]{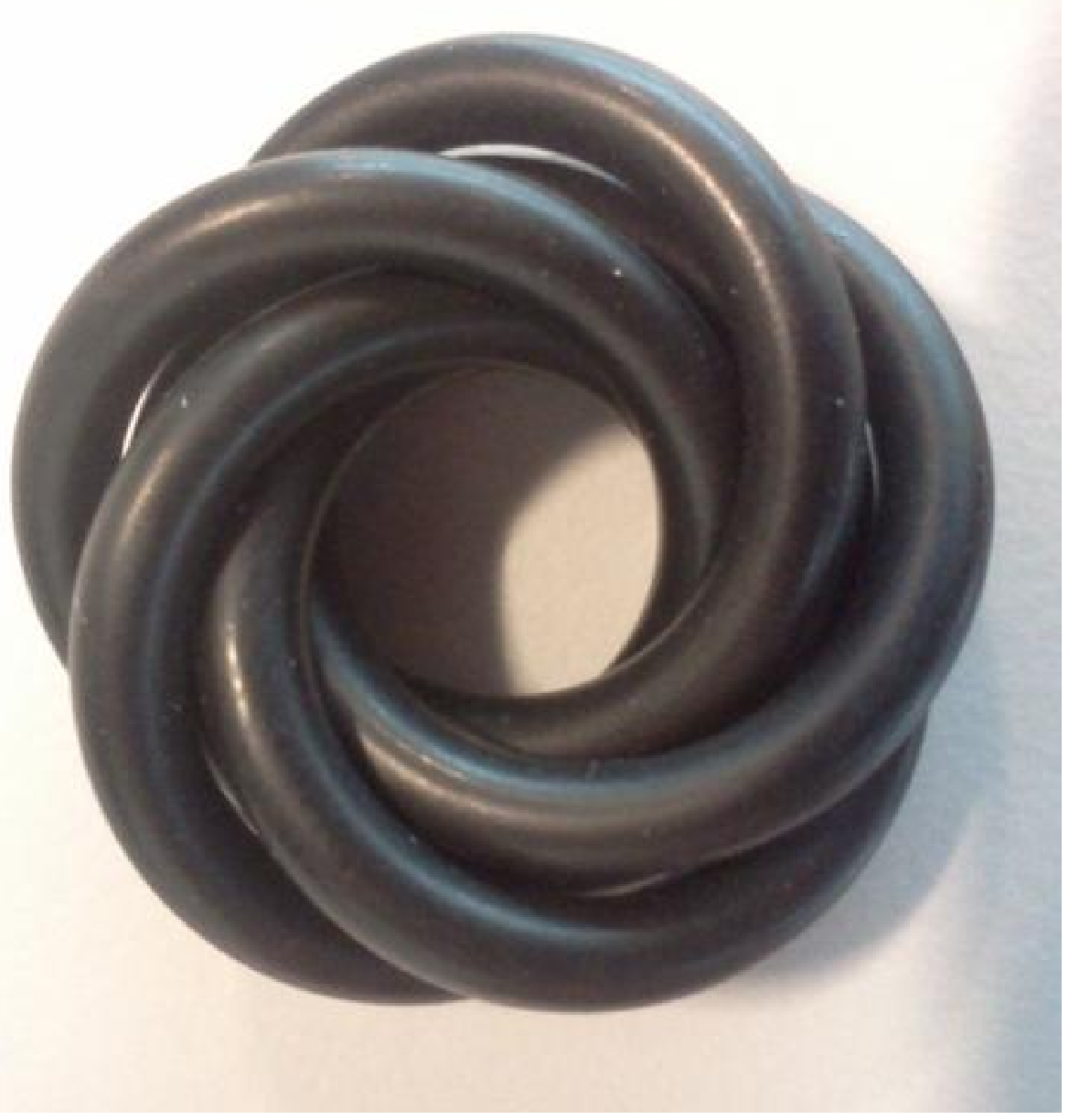} \\
\mbox{\bf (a)} & \mbox{\bf (b)}\\
\end{array}$
\caption{\label{lock}(a) Image of a  cluster of 5 interlocked elastomeric O-rings with
4-fold symmetry. (b) Image of a bundle of 6 interlocked O-rings. Such an
arrangement represents the most compact way to pack them into a
larger torus. Its cross-section includes one circle surrounded by
five toric sections.}
\end{figure}

Tori of radii ratio larger than 2 (i.e., $\gamma>2$) can be
locally packed into a dense interlocked cluster. For example,
Fig.~\ref{lock}(a) shows a cluster of 5 interlocked tori with
4-fold symmetry. Fig.~\ref{lock}(b) shows a cluster of 6
interlocked tori, which resemble a single torus. Such an
arrangement represents the most compact way to pack them into a
larger torus. Its cross-section includes one circle surrounded by
five toric sections. This leads to the possibility of constructing
dense packings of ring tori with large radii ratios possessing
self-similar properties. However, due to the complexity of the
local packing arrangements, such constructions would involve
nontrivial analytical treatments.

\section{Conclusions}


In this paper, we have reported analytical constructions of dense
packings of tori in $\mathbb{R}^3$, which are elementary examples
of multiply connected solids of genus one. Specifically, we have
obtained a family of periodic packings of unlinked, congruent tori
with a wide range of radii ratios. In this family, the packings of
tori with radii ratio $\gamma \in (0, ~1.7408\ldots)$ possess a
higher density $\phi$ than the optimal sphere packing density
$\pi/\sqrt{18}=0.7404\ldots$. In other words, all spindle tori and
the horn tori pack more densely than spheres; and even certain
ring tori can pack more densely than spheres. A maximum packing
density $\phi_{\mbox{\tiny max}} = 0.7863\ldots$ is achieved by
spindle tori of radii ratio $\gamma^*=0.8203\ldots$.

For ring tori with large radii ratios, the large void space in the
holes of the tori eventually makes them bad packers. However, one
can take advantage of the holes by considering linked or
interlocked clusters of tori to achieve higher packing densities.
For example, when two ring tori with $\gamma = 2$ are arranged in
a Hopf link dimer with central symmetry, a very dense lattice
packing can be obtained. This indicates that the organizing
principles originally formulated for simply connected convex and
non-convex solid bodies \cite{organizing} also apply to these more
complex shapes.

Finally, we would like to note that for ring tori with very large
radii ratios, more sophisticated packing arrangements could be
identified that possess much higher packing densities than the
simple periodic constructions based on the cylindrical-like
stackings of the rings. However, such complex arrangements are
difficult to find analytically. A recently developed
adaptive-shrinking-cell optimization procedure \cite{asc} could be
used to search numerically for such complex packings of tori.

\begin{acknowledgments}
R.G. was supported by the Autonomous Province of Trento and the
European Union's Seventh Framework Programme for research (COFUND
- Call 1 - Postdoc 2010). Y. J. was supported by the start-up
funds provided by Arizona State University. S.T. was supported in
part by the Division of Mathematical Sciences at the National
Science Foundation under Award No. DMS-1211087. This work was
partially supported by a grant from the Simons Foundation (Grant
No. 231015 to Salvatore Torquato).
\end{acknowledgments}


\begin{thebibliography}{13}
\bibitem{fn1}
We note that another elementary example of a multiply connected
non-convex body is a sphere in which a very thin cylinder passing
through the sphere center is removed. Such a modified sphere is topologically equivalent to
a torus, but the packing fraction of configurations of such genus-one objects
can never exceed that of the solid sphere, unlike packings of tori. This clearly illustrates the importance of
the geometry of the particle, not just the topology, in
determining its packing behavior.
\end{thebibliography}

\end{document}